\documentclass[iop]{emulateapj}
\usepackage{apjfonts}
\usepackage{graphicx}
\usepackage[caption=false,labelfont=large]{subfig}
\usepackage{color}
\usepackage{wrapfig}
\usepackage{calc}

\usepackage{epstopdf}
\usepackage{amsmath}
\def\ltsima{$\; \buildrel < \over \sim \;$}
\def\simlt{\lower.5ex\hbox{\ltsima}}
\def\gtsima{$\; \buildrel > \over \sim \;$}
\def\simgt{\lower.5ex\hbox{\gtsima}}

\usepackage{color}

\shorttitle{Throwing Icebergs at White Dwarfs}
\shortauthors{A.P. Stephan et al}

\begin{document}

\title{Throwing Icebergs at White Dwarfs}

\author{Alexander P. Stephan, Smadar Naoz\altaffilmark{1}, B. Zuckerman }

\affil{Department of Physics and Astronomy, University of California, Los Angeles, CA 90095, USA}
\altaffiltext{1}{Mani L. Bhaumik Institute for Theoretical Physics, University of California, Los Angeles, CA 90095, USA}
\email{alexpstephan@astro.ucla.edu}

\begin{abstract}
White dwarfs have atmospheres that are expected to consist nearly entirely of hydrogen and helium, since heavier elements will sink out of sight on short timescales. However, observations have revealed atmospheric pollution by heavier elements in about a quarter to a half of all white dwarfs. While most of the pollution can be accounted for with asteroidal or dwarf planetary material, recent observations indicate that larger planetary bodies, as well as icy and volatile material from Kuiper belt analog objects, are also viable sources of pollution. {The commonly accepted pollution mechanisms, namely scattering interactions between planetary bodies orbiting the white dwarfs, can hardly account for pollution by objects with large masses or long-period orbits.} Here we report on a mechanism that naturally leads to the emergence of massive body and icy and volatile material pollution. {This mechanism occurs in wide binary stellar systems, where the mass loss of the planets' host stars during post main sequence stellar evolution can trigger the Eccentric Kozai-Lidov mechanism. This mechanism leads to large eccentricity excitations, which can bring massive and long-period objects close enough to the white dwarfs to be accreted.} We find that this mechanism readily explains and is consistent with observations.
\end{abstract}


\maketitle

\section{INTRODUCTION}\label{sec:Intro}

Most stars ($\leq 8$~M$_\odot$) that have exhausted their nuclear fuel end up as white dwarfs (WDs). Over the last few decades, many WDs have been observed with spectra that show the presence of significant amounts of elements heavier than hydrogen or helium in their atmospheres \citep{Zuckerman+2003,Zuckerman+2010,Koester+2014}. These heavy elements are expected to sink rapidly to the core of WDs, which implies a recent replenishment \citep{Paquette+1986_A}. The commonly expected source for this replenishment is material from asteroidal or minor planet size bodies \citep{DebesSigurdsson2002,Jura2003,Veras+2017_B}, some of which have been observed in the process of tidal break-up \citep{Vanderburg+2015,Xu+2016}. However, some observations have suggested that planetary bodies, with sizes of the order of Mars or larger, can also contribute to WD pollution \citep{Jura+2009,Zuckerman+2011}. Furthermore, for the first time a WD (WD 1425+540) has shown signs of pollution by icy and volatile material from a Kuiper belt analog object that was initially on a very wide orbit \citep{Xu+2017}. 
Here we report on a mechanism that naturally leads to the emergence of both of these observed features{, large planetary mass and icy and volatile material pollution}.
In this mechanism we utilize the observed high binary fraction of stars \citep{Raghavan+10} that facilitates the accretion of long-period planets and Kuiper belt analog objects onto WDs.
Secular (i.e., long term coherent) gravitational perturbations exerted by a stellar companion on such objects can excite their eccentricities to extreme values, and even lead to accretion onto the host star. If these extreme eccentricity excitations take place after the host star has evolved to the WD phase, the WD can be polluted by a planet or a Kuiper belt analog object. 

Most efforts to explain the aforementioned polluted WD observations have focused on mechanisms to bring rocky asteroids onto a WD, either through scattering or secular effects \citep{Veras+2013,HPZ2016,Veras+2017,Petrovich+2017}. These can explain small mass object accretion by material that orbits the WD on relatively close orbits, as well as Fe, Mg, O, and Si signatures in the WD atmosphere \citep{Veras+2013,Farihi2016,HPZ2016,Veras2016,Veras+2017,Petrovich+2017}. However, they cannot readily account for, or have not been used to explain, accretion of planetary size objects \citep{Jura+2009,Zuckerman+2011} or icy and volatile material \citep{Xu+2017} from wide orbits. Here, we consider a potential polluter (PP), whose mass can be anywhere between several times the mass of Jupiter to the mass of a large asteroid, orbiting its host star on a relatively wide orbit that is being gravitationally perturbed by a much more distant stellar companion through the Eccentric Kozai-Lidov mechanism.


{In this paper, planetary material refers to rocky and metallic material in large amounts (i.e. from Mars-size, or larger, bodies), while icy material refers to water ice, and volatile material refers to volatile chemicals based largely on nitrogen, carbon, or sulfur. Neptune-like ice giants and Kuiper belt analog objects contain both icy and volatile material, while ice-giants also have large rocky and metallic cores. Volatile material, and thus nitrogen, is difficult to bring onto a WD as its snow line is much further from a star than for planetary or icy material. While the accretion of volatile material most noticably enriches a WD's atmosphere in carbon and nitrogen \citep{Xu+2017}, the accretion of icy material will over time accumulate hydrogen in the atmosphere (especially noticeable for helium-dominated WDs), since hydrogen always remains in the atmosphere \citep[e.g.,][]{GentileFusillo+2017}.} 

The majority of WD progenitors have a binary companion \citep{Raghavan+10} that can excite the eccentricities of PPs to extreme values \citep{Naoz2016}. This can cause some of these PPs to accrete onto the primary star during its main sequence lifetime or to be engulfed by the star as it becomes a red giant. During the Asymptotic Giant Branch (AGB) stage, stars lose mass, which causes the orbits of surviving PPs and distant stellar companions to expand. After this stage, if a companion can trigger extreme eccentricity excitations, accretion onto and pollution of WDs can occur. 

The paper is organized as follows: We begin by describing the numerical setup of our calculations and Monte Carlo simulations (Section \ref{sec:ICs}), followed by a description of the orbital parameters that lead to accretion (Section \ref{sec:Formation}). We end the paper by discussing the implications of our results and our conclusions (Section \ref{sec:dis}).

\section{Initial Conditions and Numerical Setup}\label{sec:ICs}

\subsection{Monte Carlo Simulations}\label{subsec:MC}

We perform large Monte Carlo simulations of two representative example scenarios covering different mass scales, Neptune-like planets (denoted {\bf Neptune-runs}) \citep{Howard2013} and Kuiper belt analogs (denoted {\bf Kuiper-runs}), and give a proof of concept for the proposed pollution mechanism. The initial parameters for these systems are chosen to be consistent with observed main sequence binary stars (see Table \ref{tbl:runs} for an overview of the parameters). As such we choose the primary stellar mass 
 from a Salpeter distribution \citep{Salpeter1955}, limited between $1$ and $8$~M$_\odot$. More massive stars are not expected to evolve into WDs, while less massive stars have not had enough time to evolve to become WDs over the age of the Galaxy. The mass of the companion star is chosen from a normal distribution of mass ratios consistent with observations of field binaries \citep{Duquennoy+91}.  The masses and radii of Neptune-like planets are set equal to Neptune's, while the mass and radius of Kuiper belt dwarf planet Eris is used for the Kuiper belt analog objects.

\begin{deluxetable}{c rrr}
\tablecaption{Initial Parameters before applying stability criteria\label{tbl:runs}}
\tabletypesize{\scriptsize}
\tablecolumns{4}
\tablewidth{\columnwidth}
\tablehead{
	\colhead{Parameter} & \colhead{Neptune-runs} & \colhead{Kuiper-runs} & \colhead{WD 1425+540 runs}
	}
\startdata
	{\# of runs} & 3000 & 1500 & 1500 \\
	a$_{PP,i}$ [AU] & {20-50} & 40-100 & 120-300 \\
	a$_{c,i}$ [AU] & $\gtrsim$ 200$^{D{\&}M}$  & $\gtrsim$ 400$^{D{\&}M}$ & 1120 \\
	m$_{\star,i}$ [M$_{\odot}$] & 1-8$^S$ & 1-8$^S$ & 2 \\
	m$_{PP,i}$ [M$_{\odot}$] & $5.149 \times 10^{-5}$ & $8.345 \times 10^{-9}$ &  $8.345 \times 10^{-9}$\\
	m$_{c,i}$ [M$_{\odot}$] & $0.1-8^{D{\&}M}$ & $0.1-8^{D{\&}M}$  & 0.75 \\
	e$_{pp,i}$ & 0.01 & 0-0.15 & 0-0.15 
\enddata
\tablenotetext{$D{\&}M$}{Binary separation and companion mass distributions are taken from \citet{Duquennoy+91}, determined from binary observations.}
\tablenotetext{$S$}{Primary stellar mass distribution taken from the Salpeter Initial Mass Function \citep{Salpeter1955}.}
\tablecomments{{Listed are the initial semi-major axis (a), mass (m), and eccentricity (e) distributions for Potential Polluters (subscript $PP$), primary stars (subscript $\star$), and companion stars (subscript $c$).} Unless stated differently, all parameters distributions are uniform within the given ranges. The eccentricity of the companion star's orbit is picked from a uniform distribution between 0 and 1, while i$_{tot}$ is chosen isotropically, for all three runs.}

\end{deluxetable}

\begin{figure}[!ht]
	\centering
	\includegraphics[width=\columnwidth]{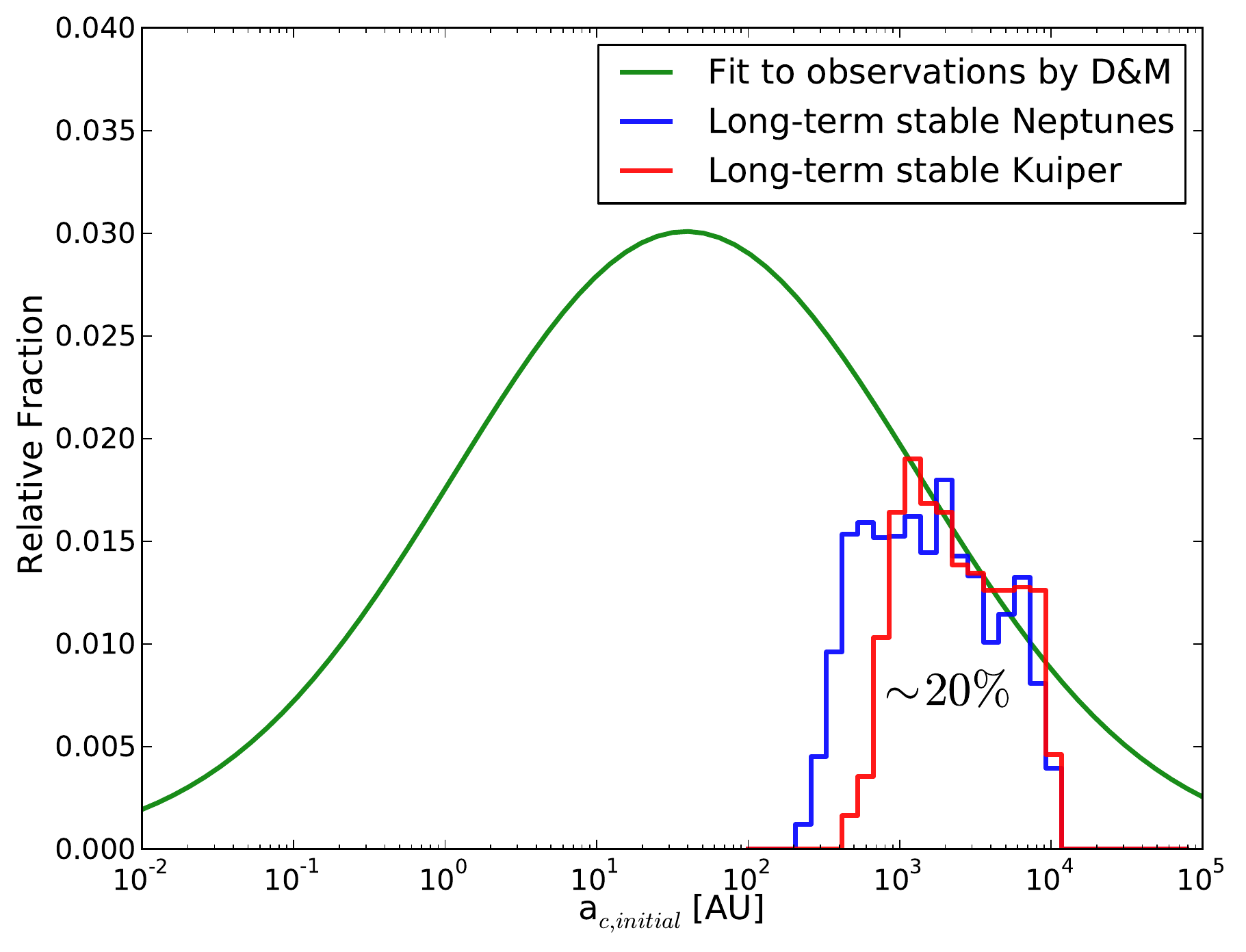}
	\caption{{\bf Initial distribution of a$_c$.} We show the initial distribution of binary star separations based on a fit to observations by \citet{Duquennoy+91} in green. Overplotted are the distributions of the companion star separation, $a_c$, for which PP orbits are long term stable. The blue histogram shows stable systems for Neptunes, the red one shows stable systems for Kuiper belt analogs. The systems in the stable distributions represent $\sim 20 \, \%$ of the total binary population. } 

	\label{fig:IC}
\end{figure}

The semi-major axis (SMA) values ($a_{pp}$) of the Neptune-like planets are chosen from a uniform distribution between $20$ and $50$~AU and set with initially very low eccentricities ($e_{pp}=0.01$), while Kuiper belt analog objects' SMA values are chosen from a uniform distribution between $40$ and $100$~AU with eccentricities chosen from a uniform distribution between $0$ and $0.15$ \citep[e.g.,][]{Trujillo+01}. Long term stability requires:
\begin{equation}
    \epsilon=\frac{a_{pp}}{a_c}\frac{e_c}{1-e_c^2}<0.1
\label{eq:stability}
\end{equation}
and ${a_{pp}}/{a_c}<0.1$ \citep[e.g.,][]{Naoz2016} so that the orbits of the PP and outer companion do not ever lead to dynamical instability and strong short-term interactions, such as, for example, scattering. This restricts the potential binary companions to wide orbits ($a_c\gtrsim200$~AU), as shown in Figure \ref{fig:IC}. We still use observational estimates from field binaries for the orbital separation \citep{Duquennoy+91}. {Note that we also reject systems with initial $a_c$ values of $\gtrsim10000$~AU, as galactic tides become strong enough to efficiently dissolve binaries of such separations.} Due to the wide binary restriction, our simulations describe only about $20 \, \%$ of the entire binary population parameter space. The companion's orbital eccentricity, $e_c$, is picked from a uniform distribution between $0$ and $1$, as long as the orbits still fulfill the long-term stability criteria. Overall we performed $3000$ {\bf Neptune-runs} and $1500$ {\bf Kuiper-runs}, {about $15 \, \%$ of which started with an initial companion star more massive than the initial primary star and which served as test systems. For these test systems no WD pollution is expected, as explained in Subsection \ref{subsec:Trigger}.}

\begin{figure*}[!ht]
	\centering
	\includegraphics[width=\linewidth]{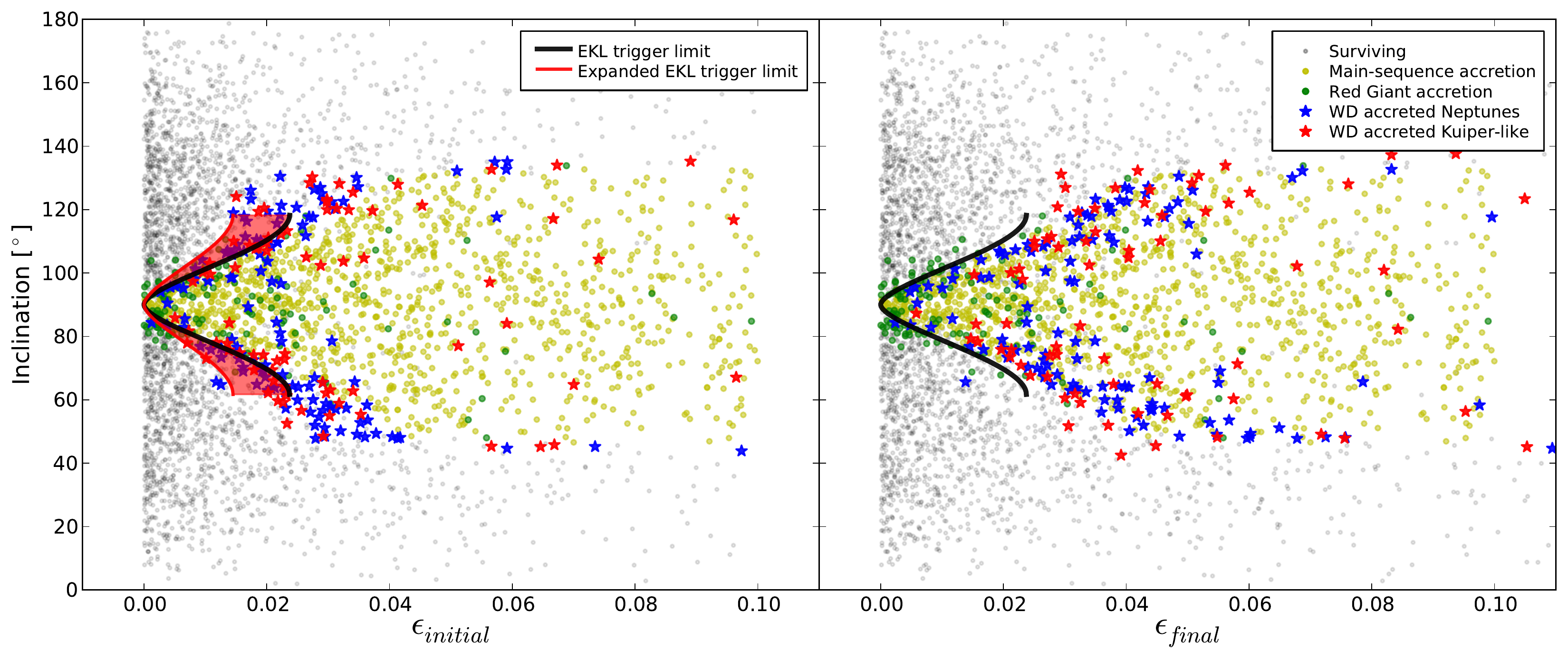}
	\caption{{\bf Octupole behavior strength within the $\epsilon$-inclination phase space.} We show the {initial} inclination {as a function of}  $\epsilon$ for our Monte Carlo simulations before (left) and after (right) mass loss has occurred. Systems that survive the full $13$~Gyr evolution are shown in grey, systems where the PP was accreted or destroyed during the main sequence are shown in yellow, and green marks PPs that were destroyed during the host star's red giant phase. Neptune-like planets and Kuiper belt analog objects that accrete onto WDs are shown as blue and red stars, respectively. The solid black line shows the theoretically predicted flip condition in the test particle case  \citep[e.g.,][]{Boaz2}, which corresponds to high eccentricity spikes, as seen, for example, on the far right side of Figures \ref{fig:1a} and \ref{fig:1b}.
	The red shaded area in the left panel shows the approximate part of the parameter space that shifts from non-EKL to EKL behavior through mass loss in an example case (m$_{\star}=2$~M$_\odot$ and m$_c=0.75$~M$_\odot$, such as WD 1425+540 and its K-dwarf companion, see \citet{Xu+2017,Wegner1981}). } 

	\label{fig:EpsIncl}
\end{figure*} 

We also perform a separate Monte Carlo simulation for the example system of WD 1425+540, which shows signs of volatile pollution suspected to stem from an accreting Kuiper belt analog object \citep{Xu+2017}. This WD has a known K-dwarf companion star \citep{Wegner1981}, and we use the simulation to determine the likelihood that our mechanism is producing the observed pollution. The WD progenitor mass is estimated to be around $2$~M$_\odot$ \citep{Xu+2017}, and the K-dwarf companion has a B-V color of about $1.29$ \citep{Zacharias+2012}, from which we estimate its mass to be approximately $0.75$~M$_\odot$. A $2$~M$_\odot$ star loses about $2/3$ of its mass before becoming a WD \citep{Hurley+00}, meaning that the total system lost about half its mass. Given that the current visual separation of the binary is $40$~arcseconds \citep{Xu+2017,Wegner1981}, which corresponds to $\sim2240$~AU, the separation before mass loss would have been $\sim1120$~AU. We use this value as the minimum possible apoapsis value for the initial binary orbital parameters, which restricts the minimum possible SMA and eccentricity value for the stellar companion. The Kuiper belt analog's SMA is uniformly chosen between $120$ and $300$~AU.
 The lower limit of $120$~AU is the roughly estimated closest distance for which a Kuiper belt analog object can retain a large amount of volatiles around a $2$~M$_\odot$ star \citep{Xu+2017}. The objects' initial eccentricities are uniformly chosen between $0$ and $0.15$. The inclination between the Kuiper belt analog's orbit and the stellar companion's orbit is chosen isotropically. After mass loss has occurred, the Kuiper belt analog objects' SMA values increase by a factor of $\sim3$, while the SMA value of the companion star only increases by a factor of $\sim2$, which leads to the increasing value of $\epsilon$. We perform $1500$ runs of possible system configurations.

\subsection{Numerical Methods and Triggering EKL}\label{subsec:Trigger}

We solve the hierarchical three-body Hamiltonian up to the octupole order of approximation and average over the orbits to obtain the hierarchical secular dynamical evolution equations, also called the Eccentric Kozai-Lidov (EKL) mechanism \citep[e.g.,][]{Naoz+11sec,Naoz2016}. In this framework the three-body system consists of an {\it inner binary} formed by the host star and a potential polluter (PP), with an initial SMA of a$_{pp}$, and which is orbitied by the stellar binary companion on a much wider orbit with SMA a$_c$, forming an {\it outer binary}.

We consider as a proof of concept two representative examples, which vary in the mass of the PPs; one is Neptune-size planets and the other is Kuiper belt analogs. We include equilibrium tidal models for the inner binary following \citet{Hut} and \citet{1998KEM}, see \citet{Naoz2016} for complete equations. We also implement general relativity precessions for the inner and outer binary \citep{Naoz+12GR}. Finally, we include radial expansion, contraction, structure changes, and mass loss due to stellar evolution for the two stars following the stellar evolution code {\tt SSE} \citep{Hurley+00}\footnote{We have tested the inclusion of post-main sequence evolution to the secular code in the past and showed that it played an important role in three-body dynamical evolution \citep[e.g.,][]{Stephan+2016,Naoz+2016}}, where we follow \citet{DD+2004} and \citet{ BarkerOgilvie2009} for the magnetic braking coefficients. 

We adopt the nominal tidal coefficent parameters for our calculations. The tidal Love number of Kuiper belt analog objects is set to $5\times10^{-5}$, since they are icy solid objects \citep{Grundy+2007,GoldreichSari2009}, for stars it is set to $0.014$ and for gas giants to $0.25$ \citep{1998KEM}. The tidal viscous evolution timescale, t$_{V}$, for stars, Neptune-like planets, and Kuiper belt analog objects is set to $1.5$~years. However, we tested different t$_{V}$ values over a range of several orders of magnitudes and found that they have no measurable influence on our results. {In particular, once a star evolves to become a WD, its small radius suppresses tidal effects on the WD, unless the orbiting planet reaches extremely close separations, at which point the planet itself will already be tidally disrupted.}

The inclusion of the octupole level of approximation leads to qualitatively different behaviors from the quadrupole level, including extreme eccentricity spikes and inclination flips from prograde to retrograde, and a generally more chaotic evolution (see \citealp{Naoz2016} for review). 
During the evolution of the system, the companion induces eccentricity and inclination oscillations on the orbit of a PP. 
In many cases the PP's eccentricity increases until it is accreted onto the evolving primary star during or before stellar expansion. However, in most cases the eccentricity excitations are not large enough to plunge the PP  onto the star before the WD phase, since the octupole effects are not strong enough. The strength of the octupole oscillations are estimated by the pre-factor of the octupole level of the Hamiltonian, $\epsilon$, which is given by Equation (\ref{eq:stability}),
see \citet{Naoz2016} for a detailed explanation. The onset of this behavior can be estimated in the $\epsilon$-inclination phase space \citep{Boaz2,Tey+13,Li+14}. This parameter space is depicted in Figure \ref{fig:EpsIncl}. where the solid black line marks the predicted onset of octupole induced inclination flips \citep[for the high inclination ($\geq 61.7^\circ$) test particle case,][]{Boaz2}.

Adiabatic (slow and uniform) mass loss in gravitationally bound two body systems of total mass $m$ leads to the expansion of the SMA $a$ according to 
\begin{equation}
    a_f=\frac{m}{m_f}a,
\end{equation}
where $f$ subscripts denote post mass loss values. In hierarchical three body systems, the mass loss in one of the inner binary members will lead to SMA expansion for both the inner binary and the outer companion. However, the total masses to consider for each SMA are different. In our case, the PP's SMA changes to
\begin{equation}
    a_{pp,f}=\frac{m_{\star}+m_{pp}}{m_{\star,f}+m_{pp}}a_{pp} \sim \frac{m_{\star}}{m_{\star,f}}a_{pp},
\end{equation}
while the companion's star SMA
 changes to
\begin{equation}
    a_{c,f}=\frac{m_{\star}+m_{pp}+m_{c}}{m_{\star,f}+m_{pp}+m_{c}}a_{c} \sim \frac{m_{\star}+m_{c}}{m_{\star,f}+m_{c}}a_{c}.
\end{equation}
 To simplify the description of the mechanism, we present a case for which the companion does not lose mass. However, we note that, throughout the rest of the calculation, we do account for any mass lost by the companion. 
With these expressions we calculate the final value of $\epsilon$, which is
\begin{equation}
 \epsilon_f=\frac{m_\star}{m_{\star,f}}\frac{m_{\star,f}+m_c}{m_{\star}+m_c} \epsilon.
 \label{eq:neweps}
\end{equation}
As m$_{\star,f}$ is always going to be smaller than $m_\star$ {and $m_c$ is larger than zero, the value of ${m_\star}/{m_{\star,f}}$ is always going to increase faster than the value of ${(m_{\star,f}+m_c)}/{(m_{\star}+m_c)}$ will decrease, due to which ${\epsilon}_f$ will always be larger than $\epsilon$.} This can move systems that were in the quadrupole dominated regime before mass loss occurred into the part of the parameter space that is octupole dominated, as shown in Figure \ref{fig:EpsIncl}. The red-shaded area in the left panel shows the estimated area of the $\epsilon$-inclination parameter space that shifts from quadrupole to octupole dominated behavior for an example case (see figure caption for details). The possibility of increasing the strength of the octupole behavior through mass loss during stellar evolution has been shown before in the context of triple stars \citep{Shappee+13}, as well as WD pollution by non-volatile material of objects smaller than Mars \citep{HPZ2016}. {If, however, the companion star is more massive than the host star and evolves first beyond the main sequence phase, the effect will be opposite. $\epsilon$ will become smaller and the strength of octupole level perturbations decreases. Likewise, once a companion star to a WD becomes a WD itself, the mass loss might move the system from the octupole-dominated regime back to the quadrupole-dominated one.}

\begin{figure*}[!p]
	\centering
	\captionsetup[subfigure]{position=top}
	\subfloat[][{\large Neptune-like Planet}]{\label{fig:1a}
		\includegraphics[width=\linewidth]{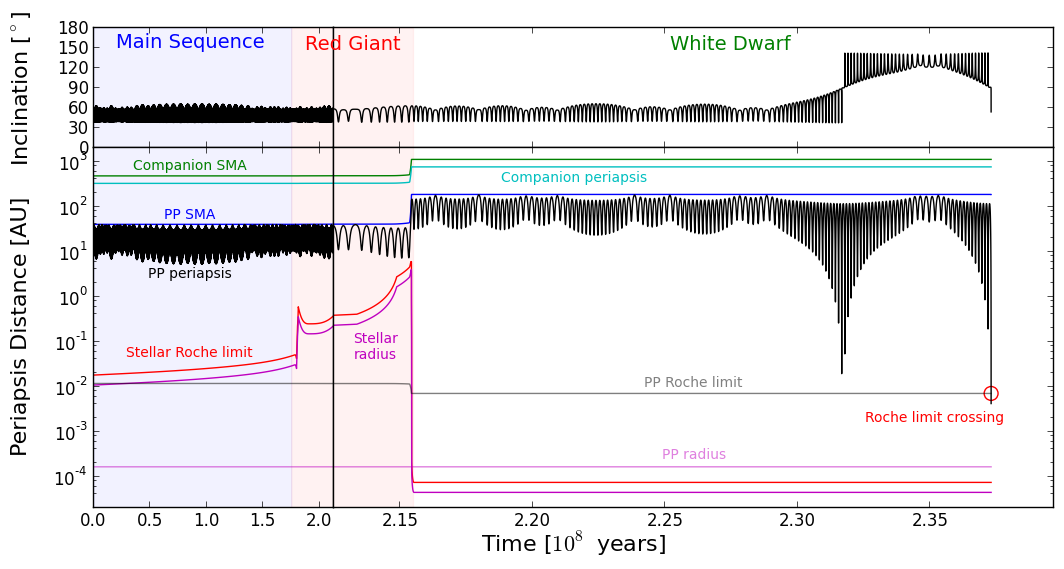}
	}
	
	\subfloat[][{\large Kuiper Belt Analog Object}]{\label{fig:1b}
		\includegraphics[width=\linewidth]{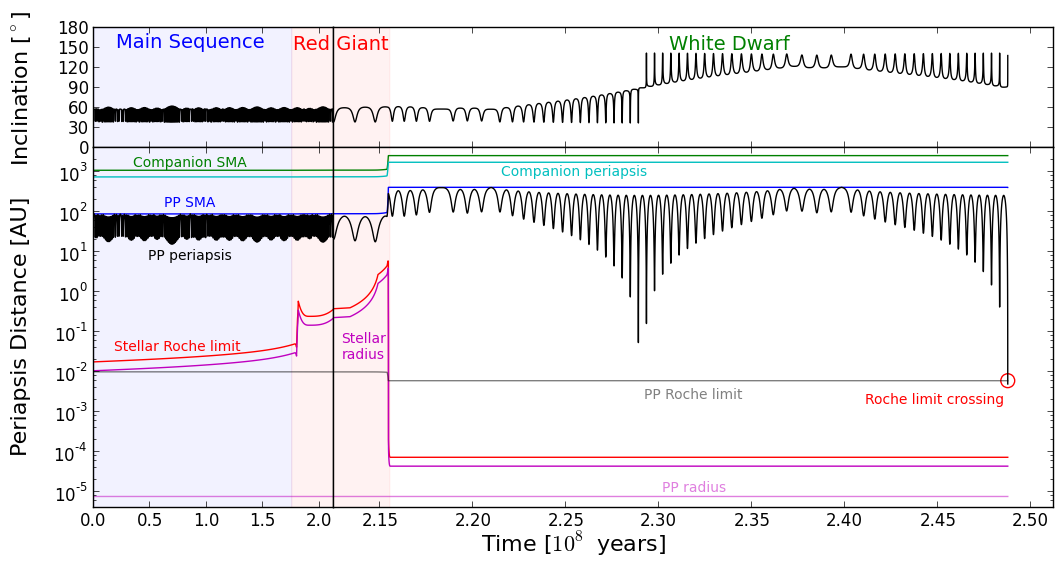}		
	}	
	\caption{{\bf The dynamical evolution of a Neptune-like planet (a) and a Kuiper belt analog object (b) in wide binary star systems.} {\bf Top parts} of each figure show the evoluton of the inclination of the potential polluter (PP). {\bf Bottom parts} show the evolution of PP periapsis distance and SMA (in black and blue, respectively), stellar companion periapsis distance and SMA (in cyan and green, respectively), host star radius and Roche limit (in magenta and red, respectively), and the PP Roche limit (in gray). {\bf Left panels} show the evolution during the fist $212.5$~Myrs of the binary systems' lifetime, while {\bf right panels} zoom in on the last (a) $\sim25$~Myrs and (b) $\sim35$~Myrs of the systems' evolution. {Note that there is no gap in the time-axis, only the scale changes at $212.5$~Myrs. The blue shaded area marks the main sequence phase, while the red shaded area approximately marks the red giant and asymptotic red giant branch phases. The white background marks the WD phase.} The initial system parameters are: (a) $m_\star=4$~M$_\odot$, $m_c=1.5$~M$_\odot$, $a_{pp}=40$~AU, $a_c=480$~AU, $e_{pp}=0.01$, $e_c=0.318$, and inclination $i_{tot}=58.265^\circ$. (b) $m_\star=4$~M$_\odot$, $m_c=1.5$~M$_\odot$, $a_{pp}=90$~AU, $a_c=1110$~AU, $e_{pp}=0.01$, $e_c=0.318$, and inclination $i_{tot}=58.265^\circ$. Inclination is defined as the angle between the inner and outer orbits' angular momenta. The value of $\epsilon$ increases by a factor of $\sim 2$ in both examples. 
	} 
\end{figure*} 

In Figures~\ref{fig:1a} and \ref{fig:1b}, we show two example evolutions where the $\epsilon$ values increase by a factor of $2$, leading to large eccentricity excitations during the WD phase (see the far right side of the lower parts of the plots). Note that periapsis distance and inclination oscillations (both in black) are fairly regular during the main sequence phase of the host stars (see blue- and red-shaded parts of the plots). The periapsis distance also does not reach extreme values, and the PPs never cross the stellar Roche limit (in red) or the stellar surface (in purple) to be destroyed or accreted. However, is shown in the parts of the plots with white background, this behavior changes after the stars have lost most of their mass during post main sequence evolution, as marked by the expanding semi-major axes of the PPs and companion stars (in blue and green, respectively, with the companion stars' periapsis distances in cyan). The eccentricities can now reach extreme peak values, at which the periapsis distances can become small enough such that the PPs cross their Roche limits (in grey, see red circles) and disintegrate around the WD, forming rings of material, which can be accreted \citep{Veras+2014b,Veras+2015}. {The stellar Roche limit is 
\begin{equation}
    R_{Roche,\star} = 1.66 \times r_\star \left(\frac{m_\star+m_{PP}}{m_\star}\right)^{1/3},
\end{equation}
while the PP's Roche limit is 
\begin{equation}
    R_{Roche,PP} = 1.66 \times r_{PP} \left(\frac{m_\star+m_{PP}}{m_{PP}}\right)^{1/3}.
\end{equation}
Here, $r_\star$ and $r_{PP}$ are the stellar and PP radius, respectively. When a PP crosses either Roche limit we halt the simulation. We assume that the PP is lost if it crosses the stellar Roche limit during the main sequence and red giant phases. If it crosses its own Roche limit during the WD phase, we assume that the PP will be accreted onto the WD.}

\section{EKL induced WD pollution }\label{sec:Formation}

\begin{wrapfigure}[37]{l}[\columnwidth+\columnsep]{12.5cm}
	\centering
	\includegraphics[width=\linewidth]{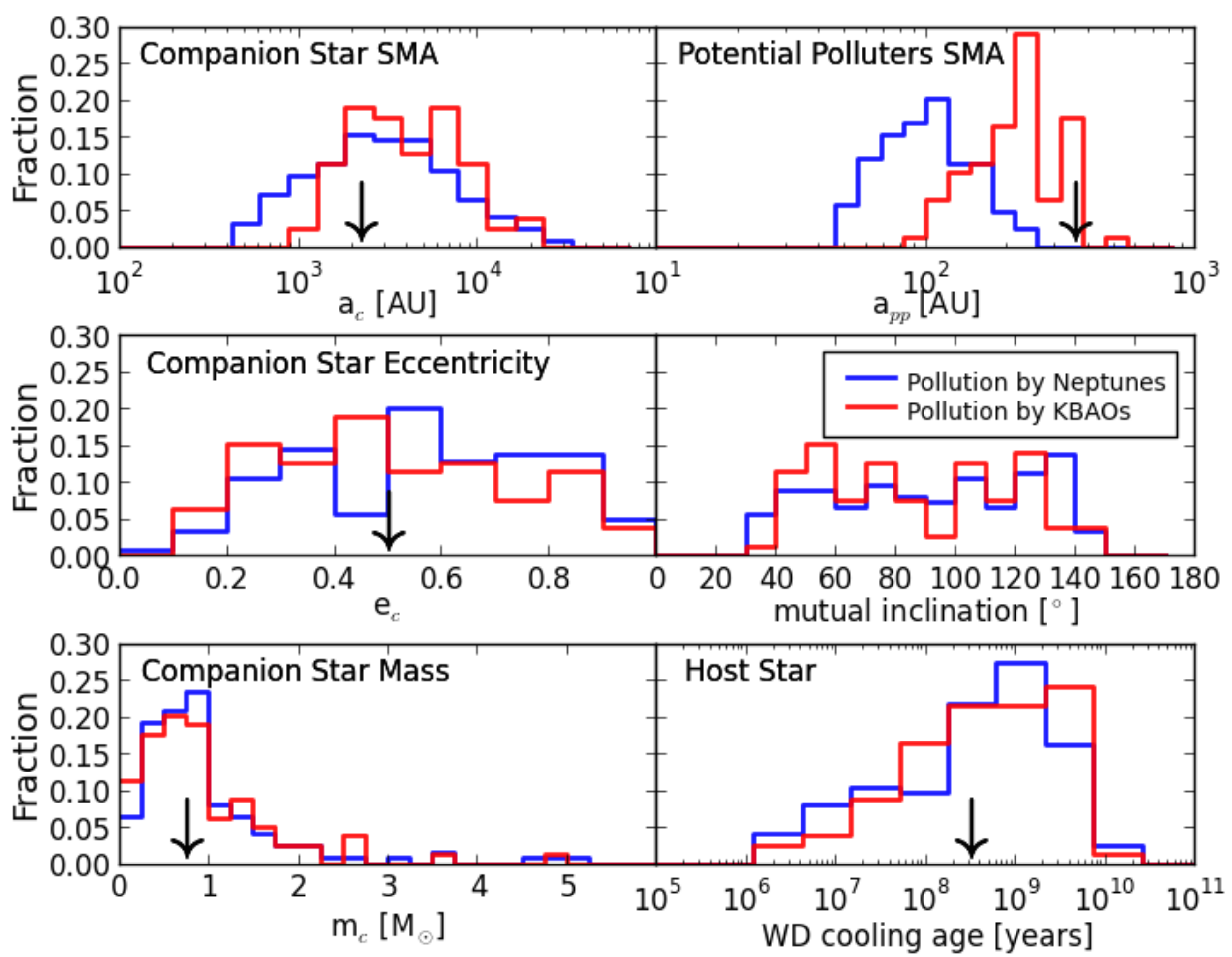}
	\caption{{\bf Orbital parameter space for polluted WDs.} We show the orbital parameter likelihood distributions for systems associated with WDs polluted by the Eccentric Kozai-Lidov mechanism, after the host star has gone through mass loss and become a WD. Shown are the parameter distributions for inducing pollution by Neptune-like planets (in blue) and Kuiper belt analog objects (KBAOs, in red). Shown are (left to right, top to bottom): SMA of the companion star and of the PP; eccentricity of the companion star and mutual inclination between inner and outer orbits' angular momenta; mass of the companion star and WD cooling age at time of accretion. The histograms are normalized such that the integral of each one is unity. The black arrows mark the known and estimated parameters for the WD 1425+540 system (see Figure~\ref{fig:Parameters2} for tighter estimation of the parameter space of this system). We note that the position of our Neptune polluters are consistent with the HR 8799 \citep{Marois+2008} planetary system.} 					
	\label{fig:Parameters}
\end{wrapfigure}

From the Kuiper and Neptune set of Monte Carlo runs we predict the orbital properties of binary stars that can lead to WD pollution through planets or Kuiper belt analog objects. Stellar companions that facilitate this mechanism are likely to have a semi-major axis on the order of a few thousand AU and are fairly eccentric ($>0.2$), as depicted in Figure~\ref{fig:Parameters}, top and middle left panels. Furthermore, the mass of the companion is most likely slightly less than a solar mass (Figure~\ref{fig:Parameters} bottom left). {Such lower mass stars have a long main sequence lifetime, which is beneficial for the EKL mechanism. Once the companion stars evolve past the main sequence phase, they lose mass, increasing the semi-major axis of the outer binary, $a_c$. As can be seen from Equation (\ref{eq:neweps}), $\epsilon$ will decrease, and further EKL evolution suppressed.} We also find that accretion can take place over a large range of WD cooling ages with a higher accretion likelihood during the first few $100$ Myrs to first few Gyrs (Figure~\ref{fig:Parameters} bottom right). {The distribution of the initial masses of host stars that lead to WD pollution was not found to differ substantially from the initial Salpeter distribution and was therefore omitted in Figure \ref{fig:Parameters}.}

We note that nearly all Neptune-like planets that were later accreted onto the WD reached periapsis distances, during their host stars' AGB phase, within a few AU of the expanding host star, as seen in the example in Figure~\ref{fig:1a}. During this phase the planet can potentially lose part of its gaseous envelope due to the strong radiation from the evolving star, which heats up the planet. This may inflate the planet's atmosphere and remove gas until only the planet's rocky core and a diminished atmosphere remain \citep[e.g.,][]{OwenWu2013,Valsecchi+14}. The body that will accrete onto the WD will therefore consist mostly of this core, not of the gaseous envelope. Furthermore, if these planets were able to retain most of their envelopes during this phase, once they plunge inwards of their Roche limit around the WD the remaining gaseous envelope will be the first part to be tidally stripped off of the planets.

In contrast to Neptune-like planets, because of their wider initial SMA, Kuiper belt analog objects that accrete during the WD phase do not typically reach periapsis distances closer than a few dozen AU during the main sequence and red giant phase, as seen in the example in Figure~\ref{fig:1b}. It is therefore highly likely that these objects will retain most or at least a significant part of their volatile material. In some cases the closest approach distance, combined with extreme radiation during the red giant phase, might trigger a comet-like behavior of the Kuiper belt analog object. {For a more detailed discussion of the thermal evolution of these objects, see \citet{JuraXu2010, JuraXu2012} and \citet{MalamudPerets2016, MalamudPerets2017}.}\vspace{37\baselineskip}

We estimate the likelihoods of accretion using binary configurations that are otherwise consistent with field binaries \citep{Duquennoy+91}, {disregarding the test systems with more massive companion than host stars, for which no WD pollution is expected and for which none occurred  (see Section \ref{sec:ICs})}. We find that $5 \, \%$ of our {\bf Neptune-runs} result in accretion of the planet onto the WD, as well as $6.5 \, \%$ of our {\bf Kuiper-runs}. {However, our {\bf Kuiper-runs} do not actually reflect the probability of WDs accreting volatile material, as they do not account for different Kuiper belt analog orbital configurations {\it per system}. A given Kuiper belt analog can be expected to contain thousands of objects with different orbital configurations, such as the solar system's Kuiper belt. For any given Kuiper belt analog in a wide binary system, at least some objects of the belt can be expected to accrete onto the WD, as long as the initial inclination of the belt to the stellar companion lies in the favorable regime (inclinations between $\sim40^\circ$ and $\sim140^\circ$, see also Figure \ref{fig:EpsIncl}). Thus, we conducted a proof-of-concept simulation for a single example system that is in the favorable part of the parameter space\footnote{We chose a fixed inclination of $70^\circ$, a WD initial progenitor mass of $1.5$~M$_\odot$, a companion star mass of $0.75$~M$_\odot$, a companion orbit SMA of $2000$ AU and orbital eccentricity of $0.5$, and SMAs and eccentricities for the Kuiper belt analog objects from uniform distributions between $40$ and $100$ AU and $0$ and $0.15$, respectively. $1000$ objects were tested, with about equal contributions to object survival, destruction during main sequence or red giant phases, and WD accretion.} and found that up to a third of the objects of a given Kuiper belt analog can over time accrete onto the WD. We estimate that about $75\,\%$ of wide binary systems systems lie in this favorable initial inclination regime, based on an inclination distribution initially uniform in cosine.} Our simulations represent about $20 \, \%$ of the stellar binary population, since we are restricted to wide binaries (Figure \ref{fig:IC}), and we adopt a $50 \, \%$ binary fraction, consistent with observations for main sequence stars \citep{Raghavan+10}. \citet{Toonen+2017} show that the binary fraction of WDs with main sequence stellar companions is lower than $50 \, \%$; they suggest that this is mostly due to stellar mergers of tight binaries, which should not influence our results. {Based on these assumptions, our simulations are applicable to about $10\,\%$ of the entire WD population.} Assuming that an average star system starts out with a Neptune-like planet (the solar system, for example, has two, Uranus and Neptune), we estimate from our results that about $1 \, \%$ of all WDs should accrete such Neptune-like planets. {Assuming that an average star system possesses a Kuiper belt analog, our results indicate that up to $\sim7.5\,\%$ of all WDs should accrete Kuiper belt analog objects. This may be consistent with observations of volatile and planetary material pollution.}

The total number of confirmed polluted WDs is on the order of $\sim200$, while the overall pollution rate for all WDs is about $25$ to $50 \, \%$ \citep[e.g.,][]{Farihi2016,Veras2016}\footnote{Note that there are $\sim1000$ polluted WD candidates known from {\it SDSS} data \citep[e.g.,][]{Zuckerman+2003, Zuckerman+2007, Zuckerman+2010, Kleinman+2013, Koester+2014, GentileFusillo+2015, Kepler+2015, Kepler+2016}.}. {However, there have been detailed abundance measurements for only about $15$ polluted WDs so far \citep[see][and references therein]{JuraYoung2014, Jura+2015, Wilson+2015, Farihi+2016, Melis+2017, Xu+2017, GentileFusillo+2017}, two of which show signs of planetary pollution in terms of composition and amount (see Section \ref{sec:Intro} for planetary pollution) \citep{Jura+2009,Zuckerman+2011}, and one with signs of volatile material pollution in the form of nitrogen \citep{Xu+2017}. While these are so far very small number statistics, these current results could imply that the planetary and volatile pollution rates are relatively high, on the order of $\sim10\,\%$, roughly consistent with our findings that $\sim 7.5 \%$ of all WDs could be polluted by Kuiper belt analog objects, while about $1 \%$ could be polluted by Neptune-like planets. More detailed abundance measurements are needed for a larger number of WDs to determine the exact occurance rates of volatile and planetary pollution.} We note here that our pollution mechanism generally operates on any objects that are large enough to avoid major orbital changes by non-gravitational forces \citep[see][for a discussion on maximum asteroid sizes sensitive to radiative forces]{HPZ2016} and are smaller than a significant fraction of the host star's mass (up to a few Jupiter masses). If any such objects commonly exist on long-period orbits in wide binary systems, they would also accrete onto WDs with a chance of $\sim1\,\%$, and could potentially be sources of planetary material pollution. Given the large uncertainty in observed planetary and volatile pollution rates, our results are consistent with the observations.

\begin{figure}[!ht]
	\centering
	\includegraphics[width=\columnwidth]{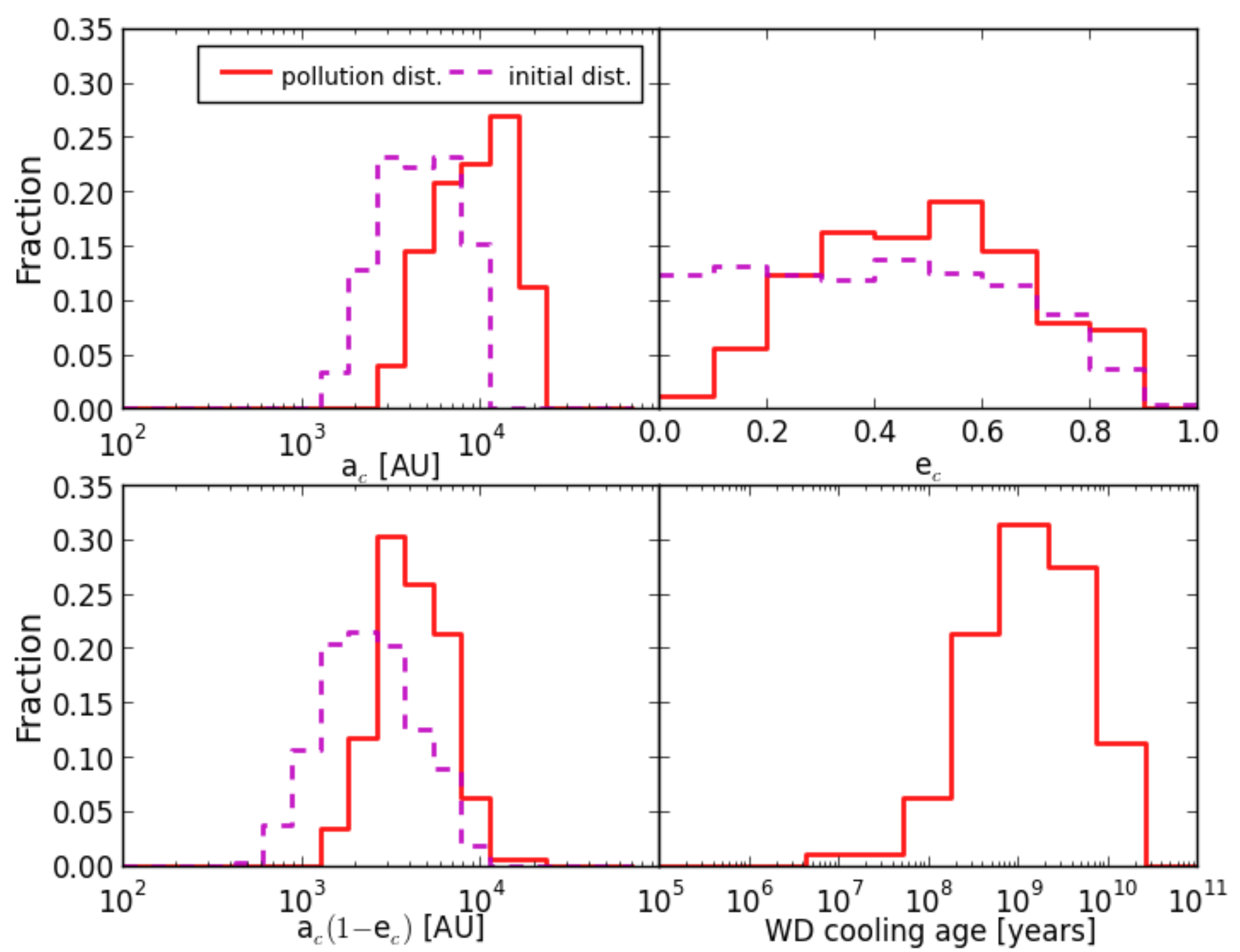}
	\caption{{\bf Orbital parameter likelihood distribution for the stellar companion of WD 1425+540.} We show the orbital parameter likelihood distributions for the K-dwarf companion to WD 1425+540 if the observed volatile pollution was caused by the Eccentric Kozai-Lidov mechanism. Shown are the initial overall tested parameter distributions (dashed, in magenta) and the post-mass loss parameter distributions that led to pollution (in red).} 
	\label{fig:Parameters2}
\end{figure}

Our additional Monte Carlo simulation for the example system of WD 1425+540 leads to accretion of icy material onto the WD for about $12.5 \, \%$ of tested configurations. From our results we estimate that the current periapsis distance of the K-dwarf companion should be about $3500$~AU, and that the mutual inclination between the Kuiper belt analog and companion star can be anywhere between $\sim30-150^\circ$ (see also Figure \ref{fig:Parameters2}). Within these loose parameter constraints our mechanism can efficiently deliver Kuiper belt analog objects onto the WD. WD 1425+540 has an age estimate of a few $100$ Myrs \citep{Xu+2017}, consistent with our estimated pollution likelihood over time (Figure \ref{fig:Parameters2}, lower right panel).

\section{Discussion and Conclusions}\label{sec:dis}

We have shown that white dwarf (WD) pollution by icy {and volatile} material from Neptune-like planets and Kuiper belt analog objects can naturally be explained through the Eccetric Kozai-Lidov (EKL) evolution in wide binary systems. The mass loss during post main sequence stellar evolution enhances the strength of EKL eccentricity excitations, which can lead to the accretion of potential polluters (PP) onto WDs (see Figures ~\ref{fig:1a} and \ref{fig:1b} as examples). 
 Systems that are more likely to facilitate this type of pollution have had an increase in $\epsilon$ value (where $\epsilon$ describes the strength of the octupole level perturbations, see Equation (\ref{eq:stability})) due to stellar mass loss during the Asymptotic Giant Branch phase. These systems occupy a specific part in the $\epsilon$-inclination parameter space (where inclination is defined as the angle between the inner and outer orbits' angular momenta), as shown in Figure~\ref{fig:EpsIncl}. 
We note that we find some PPs that were initially in the octupole favored regime were able to avoid accretion onto the WD progenitor during the main sequence and red giant phases. Accretion only took place in the WD phase for these PPs, after mass loss had occured (Figure \ref{fig:EpsIncl}).
 

\citet{Zuckerman2014} studied WD pollution in binary systems and showed that pollution should preferentely occur if the binary companion was on a wide pre-mass loss orbit of at least about $1000$~AU, which indicates that binary separations closer than that either suppress planetary system formation or prohibit long-term stability and survival into the WD phase. This is consistent with our results, where we find that the bulk of polluted WDs had initial binary separations of about $1000$~AU, as shown in Figure \ref{fig:IC}. 
This is related to the previously discussed stability criterion, which constrained our initial system architectures to be long-term stable. However, this criterion still allows for systems that are in the octupole favored regime (as depicted in Figures \ref{fig:IC} and \ref{fig:EpsIncl}). In these systems, the eccentricity excitations will drive the PPs to be accreted during the main sequence or red giant phases, before the star can become a WD. As mentioned before, only systems with mild  eccentricity excitations before mass loss can occur are able to result in WD pollution. These systems correspond to the $a_c$ distribution shown in Figure \ref{fig:Parameters}, upper left panel.

{We note here also that we ignored the effects of galactic tides on the orbital parameters of the binaries. It has been shown that galactic tides can excite eccentricities of wide stellar binaries such that their pericenter distance can get sufficiently close to scatter planetesimals onto a WD \citep{BonsorVeras2015}. However, this galactic tides mechanism acts on extremely long timescales, on the order of a few to ten Gyrs, while the EKL mechanism is most efficient for the first one or two Gyrs. In our case galactic tides would change the companion stars' eccentricity, $e_c$, which would change $\epsilon$ (see Equation (\ref{eq:stability})), and thus the strength of the octupole level of approximation. This could lead to both suppression and enhancement of EKL oscillations over time. Tides might thereby increase the efficiency of our pollution mechanism, as binaries that are outside of the appropriate inclination-$\epsilon$ parameter space could be moved inside of it. Furthermore, comparing our mechanism directly to the one in \citet{BonsorVeras2015}, it appears that galactic tides will be most relevant at polluting WDs in very wide binaries ($a_c \gtrsim 3000$~AU), while our mechanism is also effective at modestly wide separations ($a_c \gtrsim 500$ AU), for which galactic tides are weak and slow. Given that the number of binary systems drops rapidly with separation \citep[see distribution from][as plotted in Figure \ref{fig:IC}]{Duquennoy+91}, our mechanism appears to be sufficient to describe the general WD population.}

{The implications of our results for multi-planet systems broadly fall into two categories. In general, if a multi-planet system is packed tightly enough with massive enough planets, we expect EKL and large eccentricity excitations to be mostly suppressed by the planets' mutual gravitational interactions \citep{Innanen+1997}. However, if the system is not tightly packed or if the objects are not very massive, such as, for example, in a debris disk or a Kuiper belt analog, EKL excitations should still occur for each object in the disk or belt. In such systems the eccentricity and inclination changes can lead to the crossing of orbits, and potentially planet collisions or strong scattering. WD pollution should still occur for those systems. As we have discussed in Section \ref{sec:Formation}, the presence of multiple objects exhibiting a range of orbital parameters indead increases the chance of WD accretion for a given system in the favorable regime.} 

WD 1425+540, which exhibits volatile material pollution signatures \citep{Xu+2017}, has a low-mass stellar companion on a wide orbit \citep{Wegner1981}, consistent with our proposed pollution model. We are able to make additional predictions for its orbital parameters, as shown in Figure \ref{fig:Parameters2}.

We predict that WDs polluted by planets or volatile material are more likely to have a binary companion with orbital parameters consistent with the distributions in Figure \ref{fig:Parameters}, in particular large companion separations and low companion masses.
Future observations of these systems and their pollution signatures can be used to gain insights into the outer planetary and Kuiper belt analog architectures of wide stellar binary systems. {By using upcoming {\it GAIA} data releases it should be possible to find more WDs with wide binary companions due to detailed proper motion measurements. These WDs will be excellent observational targets to find more volatile pollution signatures and to investigate the long-period planetary architectures of wide binary systems.}

\acknowledgements
The authors would like to dedicate this paper to the late UCLA Professor Michael Jura for his numerous contributions to polluted white dwarf science and for inspiration. We thank the anonymous referee for fast responses and helpful comments. We also thank Siyi Xu, Dimitri Veras, and Boris G{\"a}nsike for useful discussions. SN acknowledges partial support from a Sloan Foundation Fellowship. BZ acknowledges support from a NASA grant to UCLA. Calculations for this project were performed on the UCLA cluster {\it{Hoffman2}}.


\bibliographystyle{apj}
\bibliography{Kozai2}


\end{document}